\documentstyle[preprint,revtex]{aps}
\def\gtap{\raisebox{-.4ex}{\rlap{$\sim$}} \raisebox{.4ex}{$>$}}
\def\ltap{\raisebox{-.4ex}{\rlap{$\sim$}} \raisebox{.4ex}{$<$}}
\begin{document}
\thispagestyle{empty}
\font\fortssbx=cmssbx10 scaled \magstep2
\hbox{ 
\fortssbx University of Wisconsin Madison} 
\vspace{.3in}
\hfill\vbox{\hbox{\bf MAD/PH/771}
	    \hbox{July 1993}}\par
\vspace{.2in}
\begin{title}The $W$ Boson Loop Background to $H\rightarrow ZZ$ \\
at Photon-photon Colliders
\end{title}
\author{M.~S.~Berger}
\begin{instit}
Physics Department, University of Wisconsin, Madison, WI 53706, USA\\
\end{instit}
\begin{abstract}
\baselineskip=18pt 
\nonum\section{abstract}
We have performed a complete one-loop calculation of
$\gamma \gamma \rightarrow ZZ$ in the Standard Model, including both gauge
bosons and fermions in the loop. We confirm the large irreducible continuum
background from the $W$-boson loop found by Jikia. We have included the
photon-photon luminosity, and find that the continuum background of transverse
$Z$ boson pairs prohibits finding a heavy Higgs with mass $\gtap 350$ GeV in
this decay mode.

\end{abstract}

\newpage
\section{Introduction}
The search for the mechanism of electroweak symmetry breaking
is one of the most important challenges facing
particle physics today. Detailed studies of the feasibility
of signal for Higgs bosons have been undertaken in this area
for both hadron and for $e^+e^-$ colliders. The possibility of creating a
photon-photon collider by Compton backscattering laser beams off electron beams
has attracted much attention recently.
These machines would provide collisions of photons at energies almost as
high as the parent $e^+e^-$ colliders, leading to recent interest
in the possibility of detecting
Higgs bosons at photon-photon colliders\cite{Gunion,BBC,Cheung}.
This also offers the possibility that
the Higgs-photon-photon coupling can be measured, and so provides
an indirect probe of new physics because any new charged particle that couples
to the Higgs will contribute\cite{Chanowitz}.
While the photon-photon luminosity
via the Weizsacker-Williams spectrum falls rapidly with increasing
diphoton mass, the possibility of using backscattered laser beams provides
a flat luminosity or even a luminosity which grows with energy
almost all the way to the energy of the parent $e^+e^-$ collider.

The $W$ loop contribution to $\gamma \gamma \rightarrow ZZ$ has been
calculated recently in a nonlinear gauge by Jikia\cite{Jikia} who found
a large cross section for transverse $Z$'s. We have performed an independent
calculation of this process in the Feynman gauge ($R_{\xi }$ with $\xi =1$).
We find a total cross section as well as
contributions from individual helicity modes to be in good numerical
agreement with the results of Jikia. We are also in good numerical and analytic
agreement with Glover and van der Bij who have calculated and published the
matrix elements for $gg\rightarrow ZZ$\cite{ggzz}. This result can
be immediately translated into the fermion loop
contribution for $\gamma \gamma \rightarrow ZZ$ with the appropriate coupling
replacements for the quarks and including the charged leptons.

Most studies of Higgs detection at photon-photon colliders
have neglected the irreducible continuum background of $Z$ pairs that arise
from $W$ boson and fermion loops.
The signal obtains its largest contribution from
$W$ loops, and the explicit calculation of the continuum background presented
here and in Ref.~\cite{Jikia} indicates that the contribution
to the background from almost all the
helicity modes is dominated by the $W$ boson loops as well.

The calculation required is straightforward but very lengthy. There are 188
one-loop diagrams with $W$-boson loops in the $R_{\xi }$ gauge. These are
shown in Figure 1; the mixed coupling between the photon, the charged Goldstone
boson, and the $W$ boson is present in this gauge.  The Higgs pole
appears in the diagrams in Figure 1b. We have performed the calculation with
the symbolic manipulation programs FORM and MATHEMATICA using
the tensor integral reduction algorithm of van~Oldenborgh and
Vermaseren\cite{loops}. We obtain analytic expressions for each helicity
amplitude, offering the possibility of including the full spin correlations
of the decay products of the $Z$ bosons as well as arbitrary polarization of
the incident photon beams. The result was derived for on-shell $Z$ bosons, so
it can be used for any energy above threshold
$\sqrt {s_{\gamma \gamma }}>2M_Z$.
Jikia presented the cross sections producing transverse and longitudinal $Z$
bosons. In this paper we present the cross sections of each helicity
mode of the $Z$'s separately.
We have investigated the feasibility of detecting a heavy Higgs
at a photon-photon collider by convoluting the cross section with
a realistic photon-photon luminosity
from backscattered laser beams. We present cross sections for the Higgs signal
and the continuum background. We have assumed a top quark mass of $150$ GeV
throughout this paper.

\section{Helicity Amplitudes and Cross Sections}
There are twelve independent cross sections. The contribution to
the total cross section from each helicity amplitude will be denoted
$\sigma _{\lambda _1\lambda _2\lambda _3\lambda _4}$ where $\lambda _1$ and
$\lambda _2$ are the helicities of the photons and $\lambda _3$ and
$\lambda _4$ are the helicities of the $Z$ bosons.
The cross sections for the
helicity amplitudes with equal photon helicities are shown in Figures 2a, 2b,
and 2c for Higgs masses of $M_H=300$, $500$ and $800$ GeV respectively. Only
three of the amplitudes ($++++$, $++--$, $++00$) receive
a contribution from the Higgs pole. The peaks for the heavier Higgs or
not as pronounced as the ones in
Ref.~\cite{Jikia} since we are using a top quark mass of $150$ GeV rather than
$120$ GeV for which the destructive
interference between the $W$ boson loops and the top
quark loops to the Higgs peak is more severe.
The other amplitudes are independent of
the Higgs mass. There is interference between the Higgs pole and the
continuum background that is constructive below the
Higgs peak and destructive above.

The cross sections for the
helicity amplitudes with unequal photon helicities is shown in Figures 2d.
The contributions from the other helicity modes to the cross section
are related by
\begin{eqnarray}
\sigma _{+++-}&=&\sigma _{++-+}\;, \\
\sigma _{+-++}&=&\sigma _{+---}\;, \\
\sigma _{+++0}&=&\sigma _{++0+}\;, \\
\sigma _{++-0}&=&\sigma _{++0-}\;, \\
\sigma _{+-+0}&=&\sigma _{+-0-}\;, \\
\sigma _{+-0+}&=&\sigma _{+--0}\;.
\end{eqnarray}

The cross sections $\sigma _{+-+-}$ and $\sigma _{+--+}$ are not strictly
equal, but the numerical agreement is very close. A similar statement applies
to $\sigma _{+-+0}$ and $\sigma _{+--0}$. The large cross section from the
$W$ boson loop arises in the helicity modes $\sigma _{++++}$,
$\sigma _{+-+-}$, and $\sigma _{+--+}$. These contributions reproduce the
large cross sections for transverse
$Z$ pairs at large $\sqrt{s_{\gamma \gamma}}$ first found by Jikia. For
equal photon helicities the total cross section is dominated by
$\sigma _{++++}$ in the large energy domain. Likewise,
the asymptotic value for the cross section in the case
of opposite photon helicities is given by the sum of $\sigma _{+-+-}$ and
$\sigma _{+--+}$.
These are the same helicity amplitudes
that dominate the tree level scattering
$\gamma \gamma \rightarrow W_T^+W_T^-$\cite{ppww}.
The amplitudes ${\cal M}_{++++}$, ${\cal M}_{+-+-}$, and ${\cal M}_{+--+}$
are large because they grow logarithmically with energy even at fixed
scattering angle. This is unlike
the contribution from fermion loops which approaches a constant at fixed
scattering angle.
For example, the largest helicity amplitude ${\cal M}_{++++}$ is
given asymptotically ($\sqrt{s_{\gamma \gamma}} >> M_W$) by
\begin{eqnarray}
{\cal M}_{++++}&=&8e^2g^2\cos ^2\theta _W^{}s_{\gamma \gamma}^2
\left [D(s_{\gamma \gamma},t_{\gamma \gamma})
+D(s_{\gamma \gamma},u_{\gamma \gamma})
+D(u_{\gamma \gamma},t_{\gamma \gamma})
\right ]\;,
\end{eqnarray}
where $D(s_{\gamma \gamma},t_{\gamma \gamma})$ and
$D(s_{\gamma \gamma},u_{\gamma \gamma})$ are the two straight scalar
boxes and $D(t_{\gamma \gamma},u_{\gamma \gamma})$ is the crossed scalar box
(see e.g. the first paper in Ref.~\cite{ggzz} for the functional
form of these scalar integrals in
terms of dilogarithms and elementary functions).
This amplitude is concentrated in the forward-backward directions at large
energies, and is
also logarithmically growing at large energies at fixed scattering angles.

The total cross section is the sum of the contribution from the individual
modes. Branching fractions for the subsequent decays of the $Z$ bosons have
not been included in the Figures or Tables.
For unpolarized photons there is a factor of one half from averaging
over the initial helicities.
In Figures 3 and 4 the sum of the contributions from all helicity modes is
shown for unpolarized photons with an angular cut $|\cos \theta |<0.9$ on the
$Z$ bosons. We emphasize that the angular cut is not particularly effective
at reducing the continuum background at
$\sqrt{s_{\gamma \gamma}}\approx 300-400$ GeV where the background
is fairly flat in $\cos \theta $.
The angular cut causes the cross section to begin to fall but only
in the range $\sqrt{s_{\gamma \gamma }}\; \gtap \; 700$ GeV,unlike the
process $\gamma \gamma \rightarrow W^+W^-$ where a modest cut can reduce
the transverse $W$ background by an order of magnitude\cite{Belanger}.
The prominence of the Higgs peak is reduced as the Higgs mass increases;
coupled with the rapid rise of the $TT$ background, the viability of
$\gamma \gamma \rightarrow H \rightarrow ZZ$ for detecting a heavy Higgs boson
deteriorates rapidly with increasing Higgs mass.

The peak cross section of the signal after subtracting out the
underlying background is in close agreement with the approximate form
given by the pole approximation
\begin{eqnarray}
\sigma (\gamma \gamma \rightarrow H \rightarrow ZZ)&=&
{{8\pi \Gamma (H\rightarrow \gamma \gamma )\Gamma (H\rightarrow ZZ)}
\over {(s_{\gamma \gamma }-M_H^2)^2+\Gamma _H^2M_H^2}}\;,
\end{eqnarray}
where $\Gamma _H$ is the total Higgs width.
The shape of the signal is changed
somewhat due to interference with the underlying continuum background; there is
constructive interference below the peak and destructive interference
above the peak (see Figures 2, 3, and 4). This interference tends to
be larger for the
heavier Higgs where the signal is less dominant over the background.

One sees from Figure 4 that as the Higgs mass rises from $300$ to $400$
GeV, the decreasing prominence of the Higgs peak together with the increasing
underlying continuum production of $Z$'s causes the background to become
larger than the signal. To obtain event rates one must multiply
by the $Z$ appropriate branching fractions and include the photon-photon
luminosity distribution. However, one can already deduce that
for a Higgs as heavy as $400$ GeV, one needs a large number of events
since the background is over three times the size of the peak at its maximum.

A brief comment can be made here about the Higgs bosons of extended Higgs
sectors and about supersymmetric Higgs bosons.
Extra contributions enter into the loop, e.g. the charged Higgs loops,
the squark loops, and chargino loops must be included.
These contributions, however, can be deduced from subsets of the
calculation already performed in the Standard Model. This will be the subject
of future work. In such cases,
we believe this background is still typically large.

\section{Photon-photon Colliders}
The results of the previous section were presented with respect to the
invariant mass of the photons. To understand the event rates at a realistic
photon-photon collider one must incorporate the spectrum of photons obtained
from Compton backscattering a laser beam off the electron beam.
We assume the photon beams obtained are unpolarized, thus giving a broad and
largely flat luminosity distribution up to approximately 80\% of the energy of
the $e^+e^-$ collider. We consider three such electron colliders with energies
of $E\equiv\sqrt{s_{e^+e^-}}=500$ GeV, $1$ TeV, and $1.5$ TeV.

The photon-photon luminosity is given by\cite{pplum}
\begin{eqnarray}
{{dL_{\gamma \gamma }}\over {dz}}&=&2\sqrt{\tau}k^2\int _{\tau /x_m}^{x_m}\;
{{dx}\over x}F_{\gamma/e}(x,\xi )F_{\gamma/e}(\tau /x,\xi )\;, \label{lum}
\end{eqnarray}
where $z^2=\tau ={s_{\gamma \gamma }}/{s_{e^+e^-}}$
and for unpolarized photons
\begin{eqnarray}
F_{\gamma/e}(x,\xi )&=&{1\over {D(\xi )}}\left [1-x+{1\over {1-x}}
-{{4x}\over {\xi (1-x)}}+{{4x^2}\over {\xi ^2(1-x)^2}}\right ]\;, \\
D(\xi )&=&\left (1-{4\over \xi}-{8\over {\xi ^2}}\right )\ln (1+\xi )
+{1\over 2}+{8\over \xi}-{1\over {2(1+\xi )^2}}\;,
\end{eqnarray}
with the dimensionless parameter
\begin{eqnarray}
\xi &=& {{2\sqrt{s_{e^+e^-}}\omega _o}\over {m_e^2}}\;.
\end{eqnarray}
We take the conversion coefficient $k$ to be one.
The energy of the laser beam cannot be too large, or it would be possible to
create electron-positron pairs from an interaction between the laser beam and
the backscattered photon. So we take
the dimensionless parameter $\xi $ is taken be 4.82 as usual
($\omega _o\simeq 1.26$ eV for a 500 GeV $e^+e^-$ collider)
The maximum value of the fraction
of the incident electrons's energy carried by the back-scattered photon, $x$,
is then
\begin{eqnarray}
x_m&=&{{\xi }\over {1+\xi }}\approx 0.83\;.
\end{eqnarray}
The luminosity integral in Eq.~(\ref{lum}) can be integrated to give an
analytic expression\cite{note},
\begin{eqnarray}
{1\over {2k^2z}}{{dL_{\gamma \gamma }}\over {dz}}=
{{4}\over {(1+\xi )\xi ^4D(\xi )^2}}&\Bigg [&
(-4\xi ^4-4\xi ^5+\xi^6)+(4\xi ^2+20\xi ^3+27\xi ^4+9\xi^5-4\xi ^6)z^2
\nonumber \\
&+&(-8\xi -40\xi ^2-63\xi ^3-41\xi ^4-5\xi ^5+6\xi ^6)z^4
\nonumber \\
&+&(8+32\xi +54\xi ^2+48\xi ^3+21\xi ^4-\xi ^5-4\xi ^6)z^6
\nonumber \\
&+&(-2\xi ^2-5\xi ^3-3\xi ^4+\xi ^5+\xi ^6)z^8\Bigg ]
(-\xi+z^2+\xi z^2)^{-1}(-1+z^2)^{-2} \nonumber \\
+{{1}\over {\xi ^4D(\xi )^2}}&\Bigg [&
(-8\xi ^2-8\xi ^3+2\xi^4)+(40\xi ^2+20\xi ^3-6\xi ^4)z^2
\nonumber \\
&+&(16-32\xi -48\xi ^2-20\xi ^3+7\xi ^4)z^4
\nonumber \\
&+&(16+32\xi +16\xi ^2+12\xi ^3-4\xi ^4)z^6
\nonumber \\
&+&(-4\xi ^3+\xi ^4)z^8\Bigg ]
\ln\Bigg \{{{z^2}\over {(-\xi+z^2+\xi z^2)^2}}\Bigg \}
(-1+z^2)^{-3} \nonumber \\
+{{2}\over {\xi ^2D(\xi )^2}}&&\!\!\!\!\!(4+4\xi +\xi ^2+2\xi z^2)
\ln\Bigg \{{{\xi ^2}\over {(1+\xi )^2z^2}}\Bigg \}
\;.
\end{eqnarray}
valid in the region $0<z<x_m$ and equal to zero for $z>x_m$.
The cross section is then the convolution of this luminosity  with
the helicity amplitudes
\begin{eqnarray}
d\sigma ={1\over 2}\int _{z^-}^{z^+}dz\;{{dL_{\gamma \gamma}}\over {dz}}
\Big (d\hat{\sigma }(++)+
       d\hat{\sigma }(+-)\Big )\;,
\end{eqnarray}
where $z^-$ and $z^+$ are the minimum and maximum of the energy range
to be integrated over. To increase the statistical significance of the Higgs
peak we integrate over from
$M_H-\Gamma _H < s_{\gamma \gamma} < M_H+\Gamma _H$ (we assume that
there is sufficient resolution so that
$\Gamma _{\rm res}<\Gamma _H$ even for a Higgs as light as $300$ GeV)
so that $z^-=( M_H-\Gamma _H)/\sqrt{s_{e^+e^-}}$ and
$z^+=( M_H+\Gamma _H)/\sqrt{s_{e^+e^-}}$. A factor of one-half is
included to convert these contributions from each helicity mode
for unpolarized photon beams because of averaging over the photon helicities.
On the other hand for perfectly polarized photons, one would consider only
the modes with equal photon helicities ($++\lambda _3\lambda _4$) yielding an
extra factor of 2 in the signal\cite{Gunion}.
Unfortunately this also yields a factor two
in the largest part of the $TT$ background, namely $\sigma _{++++}$.
However we expect some improvement from the unpolarized case since the
background $\sigma _{+-+-}$ and $\sigma _{+--+}$ would be reduced. Another
improvement arises from the fact that the polarized photon-photon luminosity
can peak more strongly than the unpolarized luminosity.

The cross sections in femtobarns is given in Tables 1-4 with  the angular cut
on the $Z$'s of $|\cos \theta |<1$, $0.9$, $0.8$, and $0.7$ in the
center-of-mass system. Three value of the Higgs mass are considered:
$M_H=300$, $350$, and $400$ GeV). This range of masses completely covers the
region where the Higgs signal is much larger than the continuum background
($M_H=300$ GeV) to the region where the signal is much smaller than the
background ($M_H=400$ GeV.
Both the signal and background for the 400 GeV Higgs is somewhat reduced
at the 500 GeV $e^+e^-$ collider, since part of the peak extends past the
range of the photon-photon luminosity,
$z\ltap 0.83$.

The event rates can be obtained from the figures by incorporating the
branching fractions of the $Z$ pairs to some final state. The four-jet decay
may be difficult because of the huge $\gamma \gamma \rightarrow W^+W^-$
process at tree level, so one can have one of the $Z$'s decay leptonically.
For example, consider the decay mode\cite{Cheung}
$ZZ\rightarrow q\overline{q}\ell ^+\ell ^-,\; \ell=e,\; \mu$ with a branching
fraction of $9.5\%$. With an integrated luminosity of $20 {\rm fb}^{-1}$
one has 20 signal events and 5.9 background events for a $300$ GeV Higgs at
a $500$ GeV parent $e^+e^-$ collider employing an angular cut
$|\cos \theta|<0.9$, while one has 8.6 signal events and
14 background events for a $350$ GeV Higgs boson.
For a $400$ GeV Higgs the situation is even worse, giving just 2.9 signal
events and 14 background events. There are other backgrounds that have
been considered\cite{Cheung} in this channel,
e.g. $\gamma \gamma \rightarrow \ell^+\ell^-Z$,
$\gamma \gamma \rightarrow q\overline{q}Z$, and
$\gamma \gamma \rightarrow t\overline{t}$ with subsequent decay of the top
quarks.
Using the results of Ref.~\cite{Cheung} we can estimate
that the irreducible continuum background is larger than the sum of
these  backgrounds
for $M_H\gtap 350$ GeV with a reasonable angular cut.

A higher energy collider will not improve the situation. While the reach
of the collider would be greater, less of the
photon-photon luminosity would be devoted to the region of the Higgs
peak $M_H-\Gamma _H<\sqrt{s_{\gamma \gamma }}<M_H+\Gamma _H$
for the Higgs masses of interest, $M_H<400$ GeV. For heavier Higgs boson the
continuum background is simply overwhelming.

{}From the tables, it is evident that
the angular cut is not effective at eliminating the large $TT$ background.
This is a different result than that obtained in the process
$\gamma \gamma \rightarrow W^+W^-$ for which the asymptotic cross section is
also predominantly front-back and a modest angular cut $|\cos \theta |<0.8$
reduces cross section by over an order of magnitude\cite{Belanger}.
The angular distribution of the $TT$ continuum is rather flat in the range of
interest here, $\sqrt {s_{\gamma \gamma }}\approx 300-400$ GeV. For larger
$\sqrt {s_{\gamma \gamma }}\gtap 1$ TeV
the front-back behavior becomes more evident and
the angular cut has a pronounced effect (see Figure 3).

It is possible that the ability to see the Higgs peak above the
$Z_T^{}Z_T^{}$ background can
be improved by employing more refined techniques\cite{Gunion}. Polarized
photons have been discussed above.
Cuts on the decay products of the $Z$'s can be used to
enhances the longitudinal component of the signal
over the transverse background.
These issues are subject for future study. since we have obtained the
individual helicity amplitudes, we can incorporate the decay density matrices
for the $Z$ bosons and recover the full spin correlations of the decay
products. In any case, the background is growing
very rapidly with energy and it seems unlikely that these methods could
change the basic conclusion that the Higgs is unobservable above $M_H=350$ GeV
in its $ZZ$ decay mode.

An interesting corollary to this result is that photon-photon
scattering $\gamma \gamma \rightarrow \gamma \gamma$ should show the same
behavior at very high energies, $\sqrt{s} >> M_W$. The same helicity
amplitudes should eventually grow with energy.
We are currently incorporating the decay density matrices for the $Z$ bosons.

\section{Conclusion}
We have confirmed the large $Z_T^{}Z_T^{}$ production from
photon-photon collisions at high energy. An angular cut on the $Z$ bosons
is ineffective at reducing this background. The search for
the heavy Higgs boson or for physics beyond the standard model such as
new states should take this background into consideration.
The large size of the $TT$ background casts doubt on the viability of the
process $\gamma \gamma \rightarrow ZZ$ as a ``quantum counter,'' since
any signal is probably buried beneath the Standard Model $W$ boson loop
contribution. The benefits of
polarized photon beams and cuts on the decay products of the $Z$ bosons
in detecting a heavy Higgs
including the continuum background considered here is a subject for
future study.

\acknowledgements
We wish to thank V.~Barger, G.~Bhattacharya, D.~Bowser-Chao, M.~Chanowitz,
K.~Cheung, A.~Djouadi, and C. Kao for useful
discussions. This research was supported
in part by the University of Wisconsin Research Committee with funds granted by
the Wisconsin Alumni Research Foundation, in part by the U.S.~Department of
Energy under contract no.~DE-AC02-76ER00881, and in part by the Texas National
Laboratory Research Commission under grant no.~RGFY93-221.

\newpage
{\center \begin{tabular}{|c|c|c|c|}
\hline
\multicolumn{1}{|c|}{$M_H$}
&\multicolumn{3}{|c|}{$\sqrt{s_{e^+e^-}}$} \\
\hline
\multicolumn{1}{|c|}{}
&\multicolumn{1}{|c|}{$0.5$ TeV}
&\multicolumn{1}{|c|}{$1$ TeV}
&\multicolumn{1}{|c|}{$1.5$ TeV}
\\ \hline \hline
\multicolumn{1}{|c|}{300 GeV}
&\multicolumn{1}{|c|}{11.6/3.5}
&\multicolumn{1}{|c|}{6.7/2.0}
&\multicolumn{1}{|c|}{4.0/1.2}
\\ \hline
\multicolumn{1}{|c|}{350 GeV}
&\multicolumn{1}{|c|}{4.9/8.7}
&\multicolumn{1}{|c|}{3.1/5.7}
&\multicolumn{1}{|c|}{2.0/3.6}
\\ \hline
\multicolumn{1}{|c|}{400 GeV}
&\multicolumn{1}{|c|}{1.6/8.8}
&\multicolumn{1}{|c|}{1.4/11.8}
&\multicolumn{1}{|c|}{0.90/7.7}
\\ \hline
\end{tabular}
\vskip .3in }
\begin{center}
{\bf Table 1:}  The cross section for the signal and background ($S/B$) in the
mass range $M_H-\Gamma _H<\sqrt{s_{\gamma \gamma }}<M_H+\Gamma _H$ with no
angular cut on the $Z$ bosons.
\end{center}
\vskip .3in
{\center \begin{tabular}{|c|c|c|c|}
\hline
\multicolumn{1}{|c|}{$M_H$}
&\multicolumn{3}{|c|}{$\sqrt{s_{e^+e^-}}$} \\
\hline
\multicolumn{1}{|c|}{}
&\multicolumn{1}{|c|}{$0.5$ TeV}
&\multicolumn{1}{|c|}{$1$ TeV}
&\multicolumn{1}{|c|}{$1.5$ TeV}
\\ \hline \hline
\multicolumn{1}{|c|}{300 GeV}
&\multicolumn{1}{|c|}{10.7/3.1}
&\multicolumn{1}{|c|}{6.1/1.8}
&\multicolumn{1}{|c|}{3.7/1.1}
\\ \hline
\multicolumn{1}{|c|}{350 GeV}
&\multicolumn{1}{|c|}{4.5/7.5}
&\multicolumn{1}{|c|}{2.9/4.9}
&\multicolumn{1}{|c|}{1.8/3.1}
\\ \hline
\multicolumn{1}{|c|}{400 GeV}
&\multicolumn{1}{|c|}{1.5/7.4}
&\multicolumn{1}{|c|}{1.3/10.0}
&\multicolumn{1}{|c|}{0.83/6.5}
\\ \hline
\end{tabular}
\vskip .3in }
\begin{center}
{\bf Table 2:}  The cross section for the signal and background ($S/B$) in the
mass range $M_H-\Gamma _H<\sqrt{s_{\gamma \gamma }}<M_H+\Gamma _H$ with the
angular cut $|\cos \theta |<0.9$ on the $Z$ bosons.
\end{center}
\vskip .3in
{\center \begin{tabular}{|c|c|c|c|}
\hline
\multicolumn{1}{|c|}{$M_H$}
&\multicolumn{3}{|c|}{$\sqrt{s_{e^+e^-}}$} \\
\hline
\multicolumn{1}{|c|}{}
&\multicolumn{1}{|c|}{$0.5$ TeV}
&\multicolumn{1}{|c|}{$1$ TeV}
&\multicolumn{1}{|c|}{$1.5$ TeV}
\\ \hline \hline
\multicolumn{1}{|c|}{300 GeV}
&\multicolumn{1}{|c|}{9.5/2.6}
&\multicolumn{1}{|c|}{5.5/1.5}
&\multicolumn{1}{|c|}{3.3/0.90}
\\ \hline
\multicolumn{1}{|c|}{350 GeV}
&\multicolumn{1}{|c|}{4.0/6.2}
&\multicolumn{1}{|c|}{2.6/4.0}
&\multicolumn{1}{|c|}{1.6/2.5}
\\ \hline
\multicolumn{1}{|c|}{400 GeV}
&\multicolumn{1}{|c|}{1.3/6.1}
&\multicolumn{1}{|c|}{1.1/8.1}
&\multicolumn{1}{|c|}{0.74/5.3}
\\ \hline
\end{tabular}
\vskip .3in }
\begin{center}
{\bf Table 3:}  The cross section for the signal and background ($S/B$) in the
mass range $M_H-\Gamma _H<\sqrt{s_{\gamma \gamma }}<M_H+\Gamma _H$ with the
angular cut $|\cos \theta |<0.8$ on the $Z$ bosons.
\end{center}
\vskip .3in
{\center \begin{tabular}{|c|c|c|c|}
\hline
\multicolumn{1}{|c|}{$M_H$}
&\multicolumn{3}{|c|}{$\sqrt{s_{e^+e^-}}$} \\
\hline
\multicolumn{1}{|c|}{}
&\multicolumn{1}{|c|}{$0.5$ TeV}
&\multicolumn{1}{|c|}{$1$ TeV}
&\multicolumn{1}{|c|}{$1.5$ TeV}
\\ \hline \hline
\multicolumn{1}{|c|}{300 GeV}
&\multicolumn{1}{|c|}{8.3/2.1}
&\multicolumn{1}{|c|}{4.7/1.2}
&\multicolumn{1}{|c|}{2.9/0.74}
\\ \hline
\multicolumn{1}{|c|}{350 GeV}
&\multicolumn{1}{|c|}{3.5/5.1}
&\multicolumn{1}{|c|}{2.2/3.3}
&\multicolumn{1}{|c|}{1.4/2.1}
\\ \hline
\multicolumn{1}{|c|}{400 GeV}
&\multicolumn{1}{|c|}{1.1/4.9}
&\multicolumn{1}{|c|}{1.0/6.4}
&\multicolumn{1}{|c|}{0.65/4.2}
\\ \hline
\end{tabular}
\vskip .3in }
\begin{center}
{\bf Table 4:}  The cross section for the signal and background ($S/B$) in the
mass range $M_H-\Gamma _H<\sqrt{s_{\gamma \gamma }}<M_H+\Gamma _H$ with the
angular cut $|\cos \theta |<0.7$ on the $Z$ bosons.
\end{center}
\vskip .6in

\newpage
{\Large \center Figures}
\vskip 0.5in

\noindent Fig. 1.   Generic diagrams in the $W$ boson loop contribution to
$\gamma \gamma \rightarrow ZZ$. The loops consist of all possible combinations
of $W$ bosons, Goldstone bosons and ghosts, and the number of nonzero diagrams
in each class (in a linear $R_{\xi }$ gauge) is indicated. The dashed line is
the Higgs boson.

\vskip 0.3in
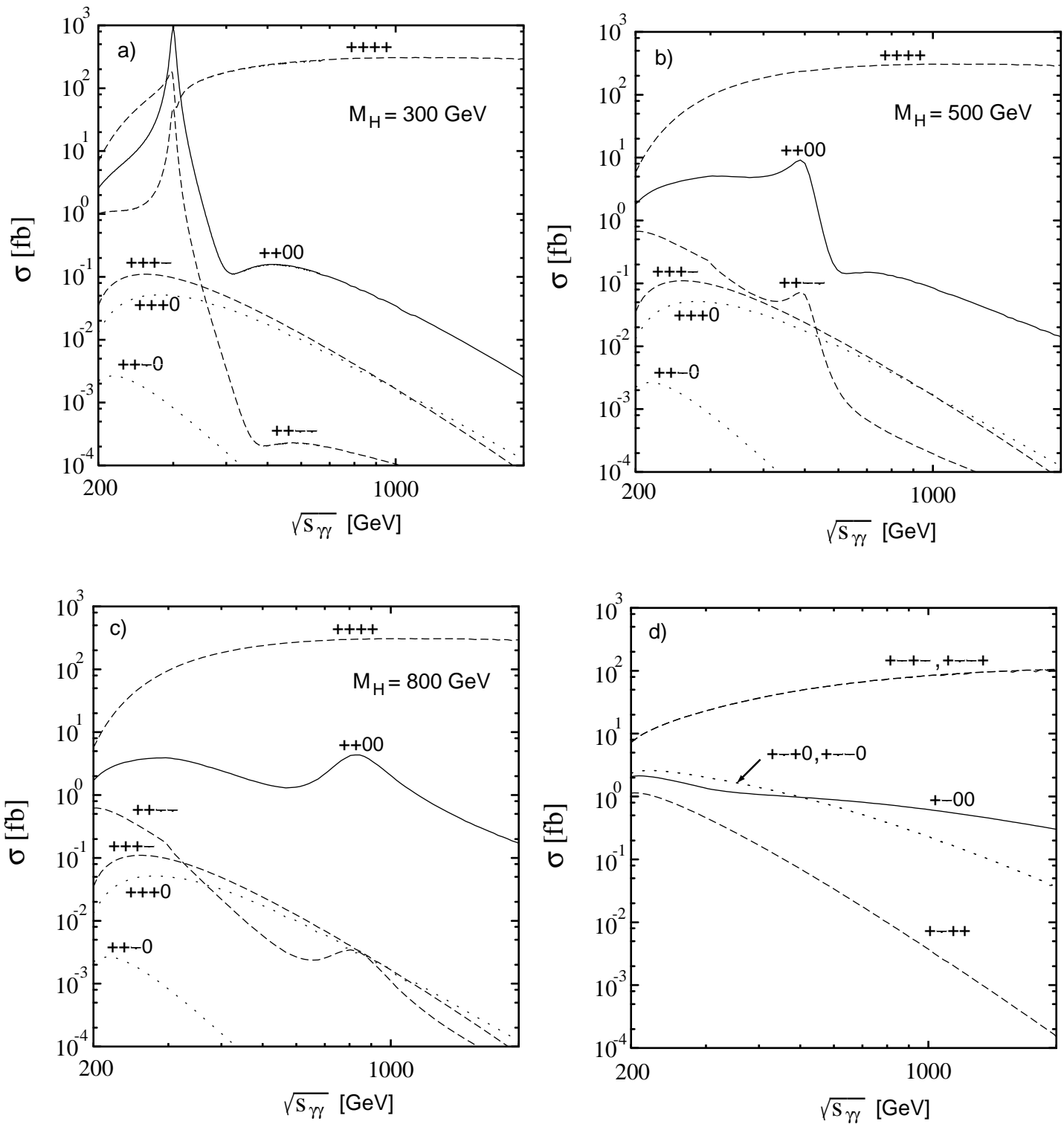
\noindent Fig. 2.   The cross sections for each independent helicity amplitude
is shown versus the diphoton mass, $\sqrt{s_{\gamma \gamma }}$. For photons
with equal helicities the Higgs mass is
$M_H=$ (a) 300 GeV, (b) 500 GeV, and (c) 800 GeV. Cross sections for
$Z_L^{}Z_L^{}$, $Z_T^{}Z_L^{}$, and $Z_T^{}Z_T^{}$ are shown wiht solid,
dotted, and dashed lines respectively. The top quark
mass is taken to be 150 GeV.

\vskip 0.3in
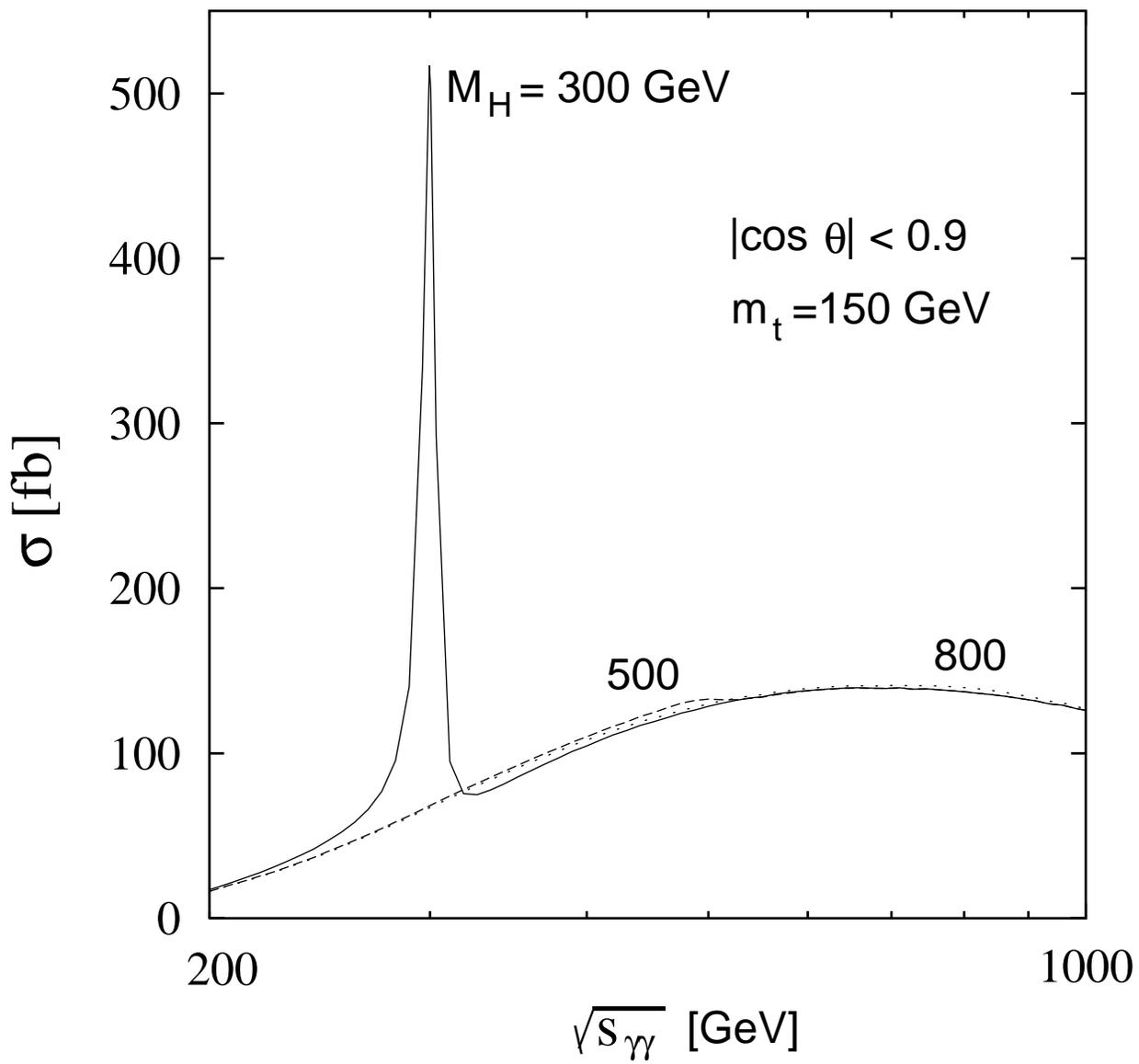
\noindent Fig. 3.   The total cross section is shown for Higgs masses of
300, 500, 800 GeV for unpolarized photons
after making an angular on the $Z$ bosons of
$|\cos \theta| < 0.9$. The top quark
mass is taken to be 150 GeV.

\vskip 0.3in
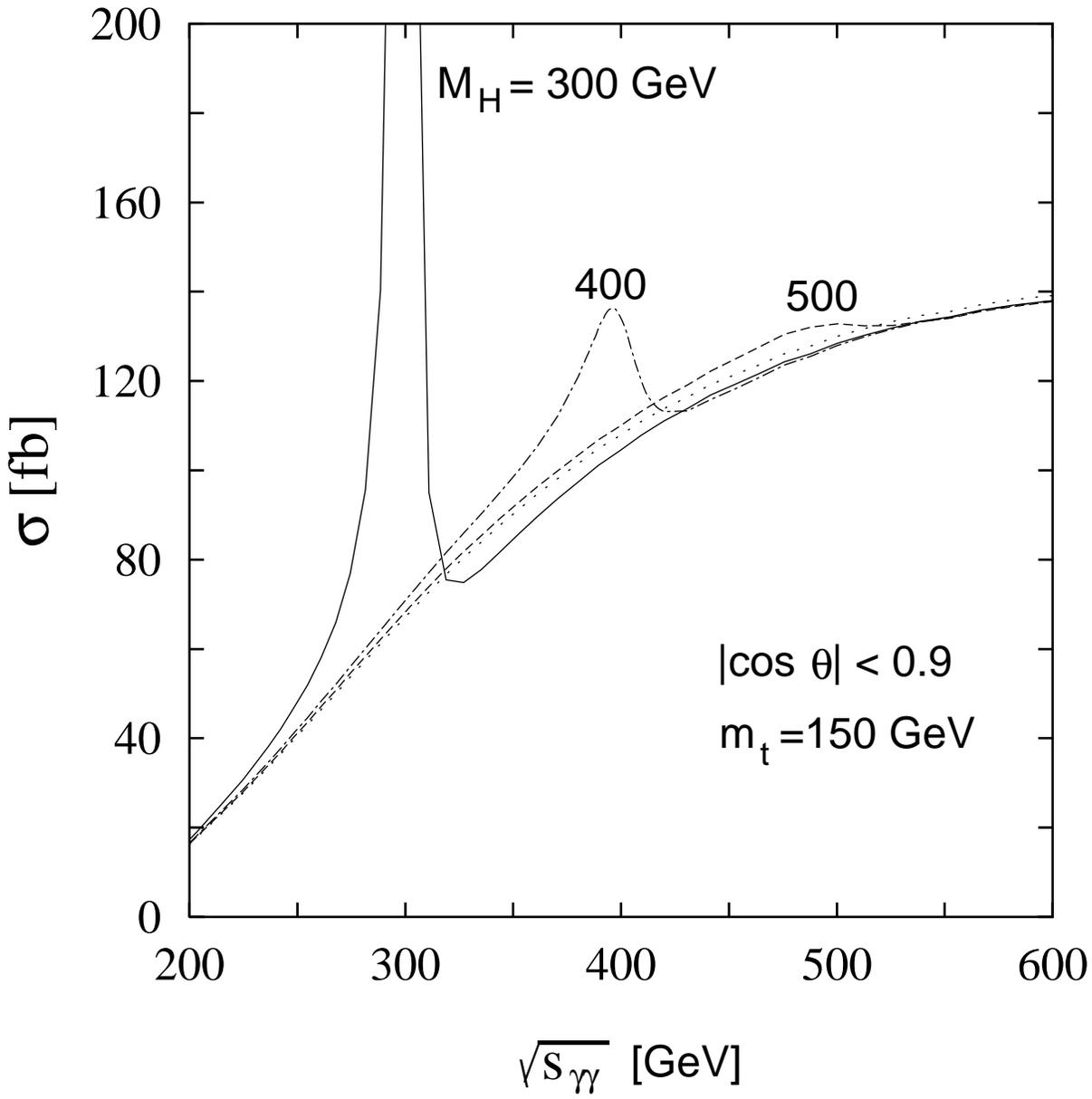
\noindent Fig. 4.   The total cross section is shown for Higgs masses of
300, 400, 500 GeV for unpolarized photons
after making an angular on the $Z$ bosons of
$|\cos \theta| < 0.9$. The dotted line is underlying background.
The top quark
mass is taken to be 150 GeV.


\begin{references}
\bibitem{note} It is
interesting to note that the luminosity integral in Eq.~(\ref{lum})
can be performed analytically even
for polarized beams.
\end{references}
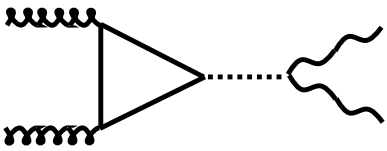
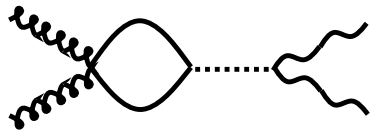
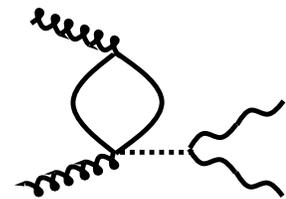
\end{document}